\documentclass{aa}
\input{psfig.tex}
\begin{document}

   \thesaurus{03(11.05.1; 11.19.2; 11.12.2,11.03.4 Coma (=Abell 1656); 
11.03.4 Cl0939+4713 (=Abell 851))} 
\title{On the universality of the type--dependent luminosity functions}

   \author{S. Andreon}


   \institute{Osservatorio Astronomico di Capodimonte, via Moiariello 16, 80131
Napoli, Italy (email: andreon@na.astro.it)}

   \date{Received ... accepted ...}

   \maketitle

   \begin{abstract} 

We find that the bright part ($M_B<-18.5$ mag) of luminosity function of
each Hubble type has the same shape in three poor clusters (Virgo, Fornax
and Centaurus) and in two rich clusters (Coma and Cl0939+4713).  The fact
that these type--dependent luminosity functions are invariant in shape
from poor to rich environments gives support to the hypothesis that they
may be universal. We also present in tabular form improved type--dependent
luminosity functions, based on a much larger sample than previous works. 


   \end{abstract}

\section{Introduction}

Luminosity functions (hereafter LFs) are powerful tools for studying the
evolution of galaxies, and a lot of attention has been devoted to their
determination and to the study of their dependence on environment (e.g.
Lugger 1989, Oegerle \& Hoessel 1989). Even more powerful is the study of
bivariate type--luminosity functions, i.e. the luminosity function of each
Hubble type (hereafter LFT), since the study of each galaxy population can
be approached separately.  For example, using the morphological
classification, Abraham et al. (1995) have shown that only a specific part
of the field population is strongly evolving. 

In a seminal paper, Binggeli et al. (1988) suggested that it is unlikely
that the shape of the LF is universal, since it is the sum of the LFTs,
and the morphological composition varies from cluster to cluster and with
respect to the intercluster field. From the the study of galaxies in the
local field, in small groups and in two poor clusters, Virgo and Fornax,
Binggeli et al. (1988) suggested the LFTs could be universal. It is clear,
however, that this statement needs to be tested in a wide variety of
environments. 

A first comparison between LFTs determined in very different environments
dates back to Binggeli (1986), who compared the $V$--band LFTs of Coma, as
measured by Thompson \& Gregory (1980), to his $B$--band LFTs of Virgo.
This comparison assumes the same color for all galaxies and that galaxies
have been classified in the same morphological scheme (whereas instead
Thompson \& Gregory (1980)'s morphological classes are not perfectly
coincident with Hubble classes).  Sandage et al. (1985), Binggeli et al.
(1988), Jerjen \& Tammann (1997, hereafter JT) and Andreon (1997a), have
compared LFTs measured in environments which are possibly not different
enough to to support the claim that LFTs are universal. 

A more robust conclusion can be drawn from a comprehensive study of
the LFTs in very different environments, where the galaxy densities differ
by 4 orders of magnitude or more. 

In Section 2 we explore the universality of the LFTs. In Section 3
we adress some objections to our claim. We build composite LFTs in
Section 4. The results are discussed in Section 5.


\begin{figure*}
\vspace{0cm}
\hbox{
\psfig{figure=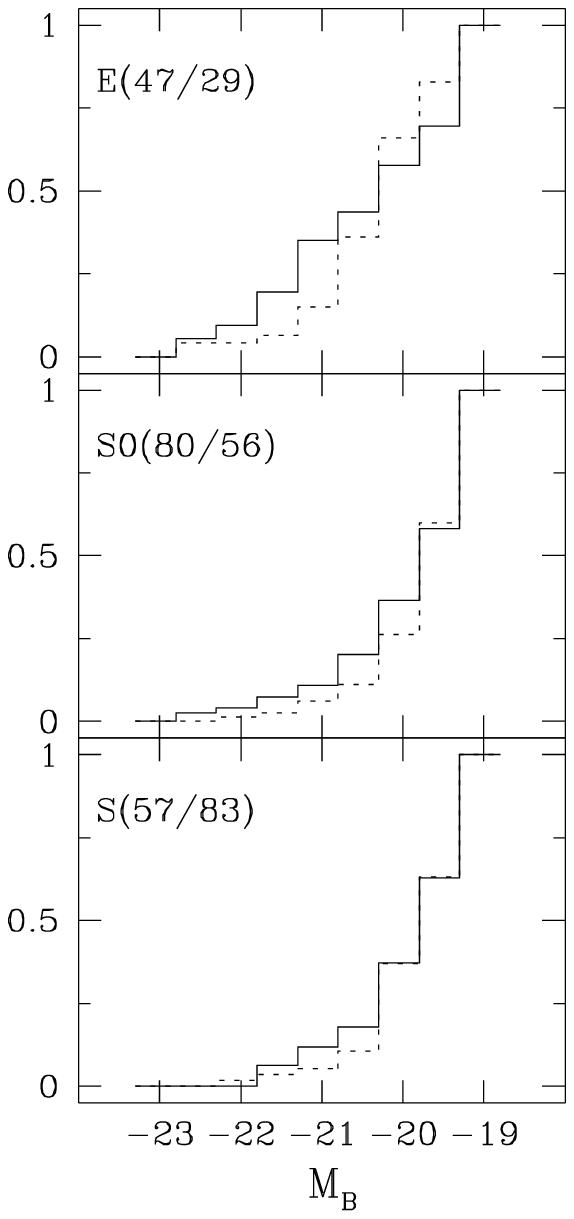,bbllx=40mm,bblly=75mm,bburx=104mm,bbury=200mm,height=9truecm,clip=}
\hspace{-0.53cm}
\psfig{figure=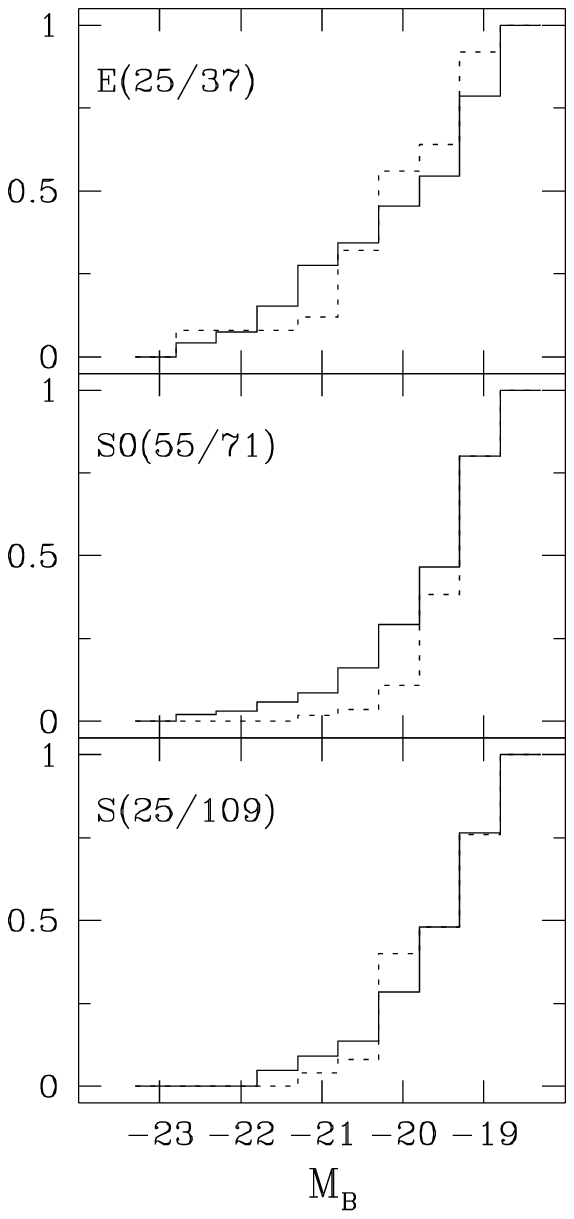,bbllx=50mm,bblly=75mm,bburx=104mm,bbury=200mm,height=9truecm,clip=}
\hspace{-0.53cm}
\psfig{figure=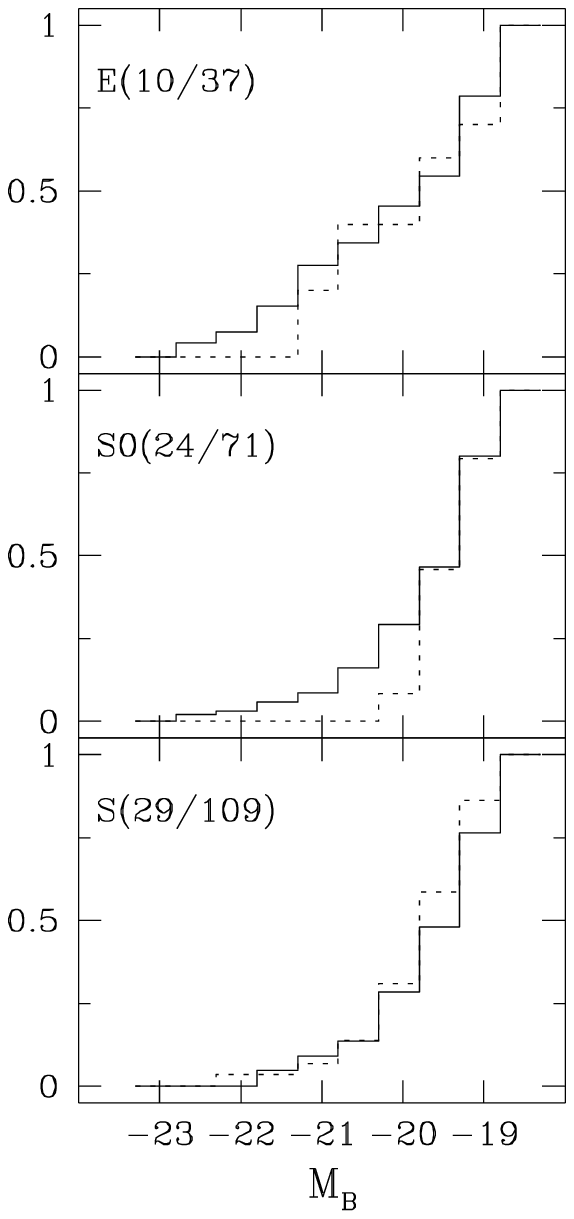,bbllx=50mm,bblly=75mm,bburx=104mm,bbury=200mm,height=9truecm,clip=}
\vbox{\hsize=60mm 
\caption{Comparison between the cumulative type--dependent luminosity
functions in poor clusters (continuos lines) with those measured in
the central region of Coma (left panels, dashed lines), in the whole Coma
cluster (central panels, dashed lines) and the Cl0939+4713  cluster
(right panels, dashed lines).  We binned our LFTs, as done by JT for their 
data, to compare our data to theirs. 
Numbers in parenthesis give
the number of galaxies of each type in the comparison clusters and in the poor
clusters, respectively}}} \end{figure*}

\bigskip
\section{Are the LFTs universal ?}

JT have recently determined FLTs in poor clusters for a sample
of galaxies including the samples of Binggeli et al. (1988) and
Sandage et al. (1985) and composed of galaxies in Virgo, Fornax and
Centaurus, grouped together in order to increase the rather poor statistics
of giant galaxies in each sample, particularly of Es and S0s. The
relative distances of these clusters are still a matter of debate and
therefore we expect that their composite LFTs will be broadened by
uncertainties in the relative distances of the clusters. 

For poor clusters we adopt the LFTs read from Figure 5 and 6 in JT. We
verified that these magnitudes were on the RC3 (de Vaucouleurs et al.
1991) $B_T$ magnitude system by comparing magnitudes of galaxies in
common. Between the magnitudes in the two systems we found a typical
scatter of $\sim 0.15-0.30$ mag and systematic differences in the range
$0$ to $0.13$ mag. Absolute magnitudes were computed by JT assuming an
{\it apparent} distance modulus for Virgo, Fornax and Centaurus of 31.7,
31.9, and 33.8 mag, respectively. 

LFTs have also been measured in two rich clusters of galaxies: Coma
(Andreon 1996) and Cl0939+4713 (Andreon et al. 1997a).  For Coma
galaxies, magnitudes are taken from Godwin et al. (1983, hereafter GMP)
and linked to the RC3 $B_T$ system comparing magnitudes of common galaxies.
We found:

$b_{GMP}=B_{RC3}+0.12 \quad \sigma=0.15$ mag.

Our choice of the distance modulus of Centaurus and the assumption that
this cluster is at rest with respect to the Hubble flow, give $H_0=65.5$
km s$^{-1}$ Mpc$^{-1}$ and a distance modulus for Coma of 35.14 mag.  For
the morphological types of Coma galaxies we have two samples at our
disposal: one complete down to $b_{GMP}=16.5$ mag with an almost complete
spatial coverage of the cluster and another complete down to $b_{GMP}=17$
mag including only galaxies in the cluster core (see Andreon et al. 1996
for details).  Rest frame $b_{GMP}$--like magnitudes of Cl0939+4713
galaxies (see Andreon et al. 1997a for details) are linked to $B_T$
assuming for them the same transformation that holds for Coma galaxies.
Morphological types are taken from Andreon et al. (1997a). We assume a
distance modulus of 41.57 mag for Cl0939+4713. 

To summarize, for all galaxies we have $J-$like magnitudes converted to
$B_T$ system, as usual for such studies, and good estimates of the
morphological types. 

The left-hand and central panels of Fig. 1 compare cumulative LFTs in poor
clusters and in the two Coma samples. Coma and the poor clusters differ by
2 orders of magnitude in central density. LFTs of the poor clusters are
statistically indistinguishable from Coma ones, down to the adopted
magnitude limits (-18.75 mag for the whole Coma cluster and -18.25 mag for
its central part). A two-side Kolmogorov--Smirnov test to reject the null
hypothesis that the compared distributions are extracted from the same
parent distribution gives at most 30 \% probability (we need 95 \% for
calling them different at the 2 $\sigma$ confidence level). This agreement
is highly satisfactory, given the scatter between adopted and RC3
magnitudes, and the poor knowledge of the relative apparent distance
moduli of Virgo, Fornax and Coma. If anything, the LF of Es in poor
clusters seems more skewed toward bright magnitudes than the Coma one
(note the excess at $M_B\sim -22$ mag and the consequent deficit at
$M_B\sim -20$ mag), but at an insignificant statistical level\footnote{A
similar excess has also been noted by Capaccioli et al. (1992) from the
comparison of the Virgo E+S0 LF with a Gaussian.}. 

Therefore, the shape of LFTs is the same not only among poor clusters, all
similar in terms of galaxy densities, but in Coma as well, where the
galaxy density is significantly different. 

The right-hand panels of Fig. 1 show a comparison of LFTs in poor clusters
and in the Cl0939+4713 cluster, which is the most distant ($z\sim0.4$,
Dressler \& Gunn 1992) ACO (Abell, Corwin \& Olowin 1990) cluster and one
of the richest known (Oemler et al. 1997). Besides this, there are
no other clusters whose LFTs have been measured in a $J$--like filter. The
distant sample is complete in absolute magnitude down to $M_B=-18.0$ mag. 

Again, the LFTs of Cl0939+4713  are statistically indistinguishable
from those of poor clusters, as it could be expected due to the
similarity of the LFTs in Coma and in Cl0939+4713 (Andreon et al. 1997a). 

The results of these comparisons give a clear indication that the shape of
LFTs, down to $M_B\sim-18.5$ mag, is invariant in clusters {\it of
different richness} (we compare the average of three richness 0 clusters
to one richness 2 cluster and one of the richest clusters), and even at
different epochs. 

\bigskip
\section{Two possible objections to the universality of the LFTs}

Previously (Andreon 1996), we noted that the $V$ band LF for S0s in
Perseus may be qualitatively different from that of their counterparts in
Coma in the $J$ band. In order to improve our comparison of LFTs in the
two clusters we recompute LFTs of Coma galaxies in $V$ using our
morphological types and photometry by Godwin \& Peach (1977).
Morphological types of Perseus galaxies have been taken from Poulain et
al. (1992) and we correct Perseus $V$ magnitudes (from Bucknell et al.
1979) for Galactic absorption using reddening maps by Burstein \& Heiles
(1984). Godwin \& Peach (1977)'s $V$ catalog of Coma is not ideal for this
comparison, since it covers a smaller area than that studied in Andreon
(1996), it is not deep enough and $V$ and $b_{GMP}$ magnitudes show a
large scatter, up to 0.75 mag. Lacking better catalogs, we adopt them, but
in this way our sample of Coma galaxies becomes smaller (67 galaxies) and
complete at a rather bright magnitude (-20.25 $M_V$ mag $\sim$ -21.25
$M_B$ mag), brighter than the magnitude at which the LF of S0s seems
different in Coma and Perseus. A statistical comparison of LFTs shows
that, in fact, they are consistent at the 50 \% confidence level, down to
our magnitude completeness limit. Given the small magnitude range involved
in the comparison, this result should not be taken as conclusive. 

\begin{figure}
\vspace{0cm}
\hspace{0cm}\psfig{figure=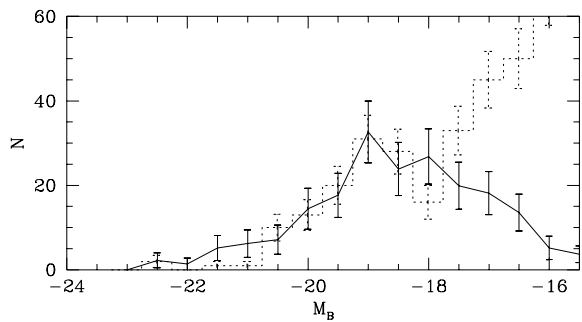,width=8cm,bbllx=20mm,bblly=65mm,bburx=165mm,bbury=150mm}
\vspace{0cm}
\caption{Real and synthetic luminosity functions of Coma (histogram
and spline, respectively), neglecting Coma dwarf galaxies and the background
contribution to the total counts, which start to be important at $M_B>-18$ mag.}
\end{figure}

The second possible objection to the universality of the LFTs concerns the
LF of Coma. It shows a bump a $b_{GMP}=16$ mag and a dip a $b_{GMP}=17$
mag, as already shown, for complete samples, by Godwin \& Peach (1977). 

Figure 2 shows that the sum of the LFTs {\it as determined in poor
clusters}, weighted by the morphological composition of Coma computed from
the data of Andreon et al. (1996, 1997b), reproduces the bump and the dip
of the Coma LF down to $M_B\sim-18$ mag. At $M_B>-18$ mag, the synthetic
LF does not match the Coma LF, because it neglects the existence of dwarfs
and of background/foreground galaxies. The rise of the LF for
$-18<M_B<-16$ can be easily modelled with a Schechter (for dwarfs)
+powerlaw (for background) functions given the small magnitude range and
the large number of free parameters for the functions. 

The maximum of the bump in the Coma LF matches well the synthetic LF.  The
bump is not sharper or shallower in the real LF than in the synthetic one.
A slight excess at $M_B\sim-22$ mag is present in the synthetic LF but
the significance is null, due to the large errorbars.  The observed dip at
$M_B\sim-18$ mag is less than 1 $\sigma$ deeper than the one expected, giving
a null statistical significance to this difference.  This is at variance
with the conclusions of Biviano et al. (1995) and Lobo (1997), but their
studies were based on less accurate comparisons. The qualitative agreement
of the shape of the $B$ LFTs in Coma and Virgo claimed by Andreon (1996)
is thus confirmed on a quantitative basis. 

\section{Improved LFTs}

\begin{figure}
\vspace{0cm}
\hspace{0cm}
\psfig{figure=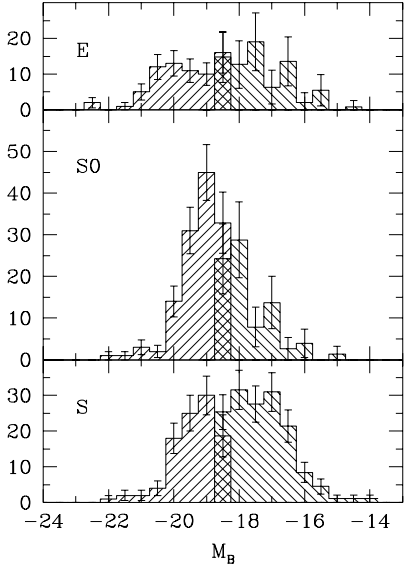,width=5.8cm,bbllx=30mm,bblly=115mm,bburx=105mm,bbury=210mm}
\vspace{0cm}
\caption{The luminosity function of Es, S0s and Ss. The two shadings 
highlight what part of the LFTs are built using rich and poor clusters
(bright and faint parts of the LFTs, respectively).
The numbers in ordinate are the observed numbers of galaxies per
half--magnitude bin, down to $M_B=-18.75$ mag in our composite sample.
For fainter magnitudes, see text.}
\end{figure}

Having shown that LFTs in different clusters are compatible among
themselves, we can now add them up. The number of early-type galaxies at
bright magnitudes in the newly added clusters (Coma and Cl0939+4713) is
more than twice those previously present in the sample. 

At bright magnitudes, because of larger photometric errors
in poor clusters, poor statistics and the open question of their
relative distances, we privileged the Coma and Cl0939+4713 LFTs
determinations, whereas at faint magnitudes we adopted LFTs in poor
clusters, since data for clusters with well known distance moduli are
missing. To be more specific, down to $M_B=-18.75$ mag, we add the entire
Coma cluster data to Cl0939+4713. For $M_B>-18.25$ mag we adopt the poor
cluster LFTs, after normalizing the number of galaxies brighter
than $M_B=-18.75$ to the one of the Coma and Cl0939+4713 clusters
together. For the half-magnitude bin centered on $M_B=-18.50$ mag we add
the samples of galaxies in Cl0939+4713, in the central part of Coma and in
poor clusters, after normalization, as above.


\begin{table}
\caption{Improved LFTs}
\begin{tabular}{cllllll}
\hline
   B& $n_E$& $\varepsilon_{n_E}$& $n_{S0}$& $\varepsilon_{n_{S0}}$& $n_S$&
$\varepsilon_{n_S}$ \\ \hline
 -23.0& ~0& 0& ~0& 0& ~0& 0 \\
 -22.5& ~2& 1.4& ~0& 0& ~0& 0 \\
 -22.0& ~0& 0& ~1& 1& ~1& 1 \\
 -21.5& ~1& 1& ~1& 1& ~2& 1.4 \\
 -21.0& ~5& 2.2& ~3& 1.7& ~2& 1.4 \\ 
 -20.5& 12& 3.5& ~2& 1.4& ~4& 2 \\
 -20.0& 13& 3.6& 14& 3.7& 18& 4.2 \\
 -19.5& 11& 3.3& 31& 5.6& 25& 5 \\
 -19.0& 10& 3.1& 45& 6.7& 30& 5.5 \\
 -18.5&	15.5& 4.4& 29.1& 5.5& 22.6& 3.8 \\
 -18.0& 12.7& 6.6& 28.8& 9.1& 31.5& 5.5 \\
 -17.5& 19.0& 8.1& ~7.9& 4.8& 27.6& 5.1 \\ 
 -17.0& ~6.3& 4.7& 13.7& 6.3& 31.0& 5.4 \\
 -16.5& 13.5& 6.9& ~2.6& 2.8& 21.4& 4.5 \\
 -16.0& ~2.1& 2.7& ~3.9& 3.4& ~8.4& 2.8 \\
 -15.5& ~5.5& 4.4& ~0& 0& ~4.5& 2.1 \\
 -15.0& ~0& 0& ~1.3& 1.9& ~1.1& 1.0 \\
 -14.5& ~0.8& 1.7& ~0& 0& ~1.1& 1.0 \\
 -14.0& ~0& 0& ~0& 0& ~1.1& 1.0 \\
\hline
\end{tabular}
\end{table}

The LFTs are plotted in Fig. 3, and the data are listed in Table 1. The
shape of the LFTs computed in JT is confirmed (and therefore their
comments hold for our LFTs too), except that the skewness of the E LF
found by JT is now much reduced for our much larger E sample. Es, S0s, and
Ss have LFTs which largely overlap in luminosity. The E LF shows a
distribution broader than the S0s. It is for this reason that, in samples
complete down to intermediate magnitudes ($M_B\sim-18$ mag), Es appear to
be brighter than S0s (e.g. Binggeli et al., 1988; Andreon 1996). S0s have
the narrowest distribution in luminosity, followed by Ss. 

The presented LFTs provide an indispensable ingredient for computing
synthetic galaxy counts, expected redshift distributions, the
observability of galaxies in voids and many other quantities which involve
the break down of the sample in populations with less heterogeneous
observational properties (k-corrections, visibility, etc.) or the
normalization at zero redshift of type-dependent quantities. 

\section{Conclusions}

We found that the bright part ($M_B<-18.5$ mag, roughly $M^*+3$) of the
LFTs is the same in three poor clusters and in two rich clusters, one of
which (Cl0939+4713) is the most distant of the Abell, Corwin, \& Olowin
(1990) catalog and one of the richest known clusters, and the other (Coma)
is the prototype of rich clusters (Jones \& Forman 1984, Sarazin 1986) and
is perhaps the best studied one. 

We have thus verified the invariance of the shape of the LFTs in a large
range of environments from as poor as the local field, where the
measured LFTs are compatibles with the ones in poor clusters (Binggeli et
al. 1988), to those of the Coma cluster and Cl0939+4713 cores (this work),
several orders of magnitude denser than the local field.  We have also
verified that the observed luminosity function of Coma is equal to
the synthetic computed assuming universal LFTs and the observed
Coma morphological composition, down to $M_B\sim-18.5$ mag. 

This suggests, in general, that differences in the LF of clusters are more
likely due to differences in morphological compositions rather than in
environment, and that the shape of LFs of each giant galaxy type (Es,
S0s, Ss) is universal. This conclusion enlarges on a previous result by
Binggeli et al. (1988) which was based on a smaller range of environments,
and confirms the results of a less direct comparison between Coma and
Virgo LFTs by Binggeli (1986). 

The over--simplified approach of studying the evolution of galaxies at
different look--back times or in different environments by comparing their
LFs or the characteristic magnitudes of the Schechter (1976) function fit
to their LF, should now be abandoned in favor of an approach that
considers the possibility that the environments being compared have
different morphological mixtures. 

Improved LFTs are given, for a much larger sample of giant galaxies than
previous works. They are quite useful for many statistical and
theoretical studied.

\begin{acknowledgements} I warmly thank E. Davoust, G. Longo and R.  de
Propris for wise suggestions that greatly helped to improve the
presentation of the results. M. Capaccioli's attentive reading of an early
version of this paper is gratefully acknowledged.  I thank the referee,
Dr. B. Binggeli, for his careful work and for suggesting an interesting
reference. \end{acknowledgements}

\end{document}